\documentclass[prb,twocolumn,superscriptaddress,showpacs,floats,floatfix]{revtex4-1}
\usepackage{dcolumn,amsmath,latexsym,graphicx,amssymb}
\usepackage[usenames,dvipsnames]{color} \usepackage{multirow}
\usepackage{array}
\usepackage{xcolor}

\RequirePackage{ifpdf}

\usepackage[usenames,dvipsnames]{color} 
\definecolor{Mygreen}{HTML}{036400}
\definecolor{Myred}{HTML}{FE0000}
\definecolor{Myviolet}{HTML}{b37bf7}
\definecolor{Myblue}{HTML}{6698FF}
\definecolor{Myorange}{HTML}{FFCA7D}
\begin{document}

\preprint{APS/123-QED}

\title{Unveiling the Signature of Adsorbed Organic Solvents for Molecular Electronics through  STM Approaches}

\author{Tamara de Ara}
\affiliation{Departamento de F\'\i sica Aplicada and Unidad asociada CSIC, Universidad de Alicante, Campus de San Vicente del Raspeig, E-03690 Alicante, Spain.}

\author{Carlos  Sabater}
\email{Corresponding author: carlos.sabater@ua.es}
\affiliation{Departamento de F\'\i sica Aplicada and Unidad asociada CSIC, Universidad de Alicante, Campus de San Vicente del Raspeig, E-03690 Alicante, Spain.}

\author{Carla Borja-Espinosa}
\affiliation{Departamento de F\'\i sica Aplicada and Unidad asociada CSIC, Universidad de Alicante, Campus de San Vicente del Raspeig, E-03690 Alicante, Spain.}
\affiliation{Yachay Tech University, School of Physical Sciences and Nanotechnology, 100119-Urcuqu\'\i, Ecuador.}

\author{Patricia Ferrer-Alcaraz}
\affiliation{Departamento de F\'\i sica Aplicada and Unidad asociada CSIC, Universidad de Alicante, Campus de San Vicente del Raspeig, E-03690 Alicante, Spain.}

\author{Bianca C. Baciu}
\affiliation{Departamento de Qu\'\i mica Org\'anica an  and Instituto Universitario de S\'\i ntesis Org\'anica,  Universidad de Alicante, Campus de San Vicente del Raspeig, E-03690, Alicante, Spain.}

\author{Albert Guijarro}
\affiliation{Departamento de Qu\'\i mica Org\'anica an  and Instituto Universitario de S\'\i ntesis Org\'anica,  Universidad de Alicante, Campus de San Vicente del Raspeig, E-03690, Alicante, Spain.}

\author{Carlos Untiedt}
\affiliation{Departamento de F\'\i sica Aplicada and Unidad asociada CSIC, Universidad de Alicante, Campus de San Vicente del Raspeig, E-03690 Alicante, Spain.}

\date{\today}

\begin{abstract}
After evaporation of the organic solvents, benzene, toluene, and cyclohexane on gold substrates, Scanning Tunneling Microscope (STM)  shows the presence of a remaining adsorbed layer. The different solvent molecules were individually observed at ambient conditions,
and their electronic transport properties  characterized through the STM in the Break Junction approach. 
The combination of both techniques reveals, on one hand, that solvents are not fully evaporated over the gold electrode and, secondly, determines the role of the electronic transport of the solvents in molecular electronics.
\end{abstract}
\maketitle

\section{Introduction}
Organic solvents such as benzene, toluene or cyclohexane are commonly used for the preparation of clean surfaces or for the deposition of other molecules \cite{TolueneHerre2008}. These apolar solvents are used under the believe that they can be easily removed through thermal evaporation, their contribution during the electrical transport measurement experiments should be negligible \cite{TolueneHerre2008,BJToluene} or that these may be distinguishable from the target molecules in the solution\cite{ElkeSolvents2016}. The understanding of the role of the solvents in the study of single molecular transport should be improved, despite all the advances of the area during the last decades \cite{Smit2002, Venkararaman2013, Jan2019, Cuevasbook, Lambertbook}.

\begin{figure}
\centering
\includegraphics[width=1.00\linewidth]{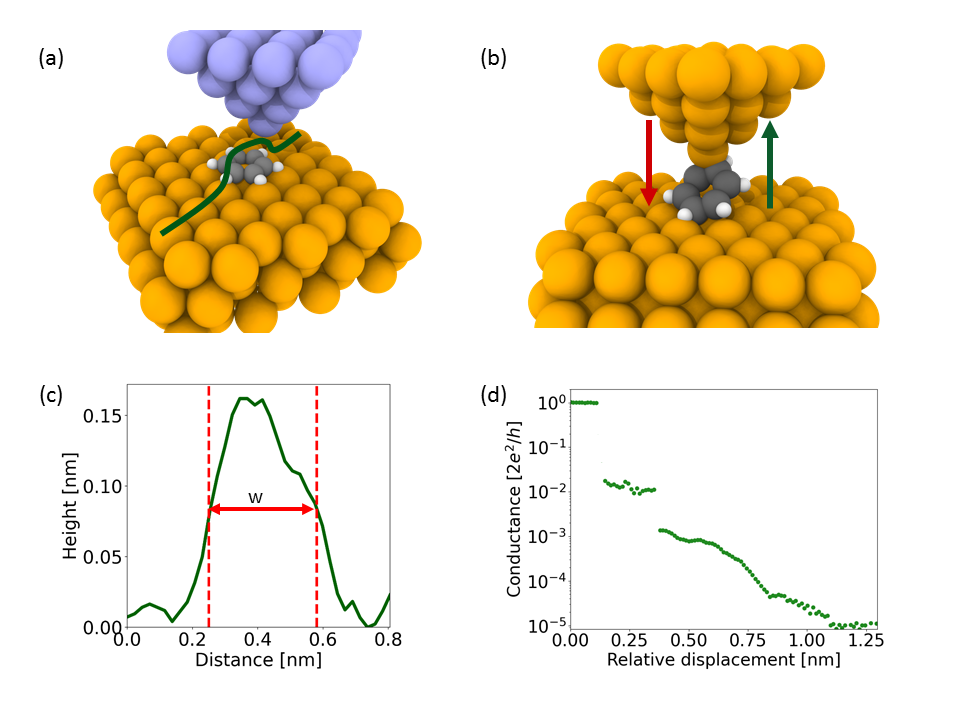}
\caption{Schematic illustration of (a) STM and (b) STM-BJ set up. Yellow and purple spheres correspond to the electrode atoms,  Au and Pt-Ir respectively. Benzene is represented with grey and white spheres, as  C and H atoms respectively. (c) Height profile of a single benzene molecule. (d) Evolution of conductance while breaking a benzene molecular junction.}
\label{fig:SchemeTechniques}
\end{figure}

In this respect, the single metal-molecule-metal junction constitutes the basic building block for molecular electronics. Many different molecules at different environmental conditions, from cryogenic\cite{Smit2002, Tal2009,SabaterOP3,MMayor2014STMBJ,Lukas2017STMBJ} to ambient conditions\cite{Xu2003,Herre2021Petide,Liu2020,Fujii2020,Zheng2018}, have been studied through the years. 
Contrary to the low temperature strategy, the study of molecular junctions at ambient conditions normally has made use of functional groups \cite{AnchoringGroups} attached to the molecule periphery, in order to help in the anchoring to the electrodes\cite{Zotti2010}. Usually a thiol \cite{thiolVenkataraman, Reedbenzenedithiol, thiolatedMolecules, Alkanedithiol,Zheng2018} group is used to enhance and strengthen the affinity. 
A paradigmatic example of these differences in strategies is the case of the study of the benzene molecule (C$_6$H$_6$). While at low temperatures different studies have been performed on the bare molecule \cite{Tal2008,benzeneWiredConductance,Tal2009}, for ambient conditions the molecule has been  functionalized, as it was commonly believed that otherwise the molecule would not remain anchored to the electrodes\cite{Reedbenzenedithiol, thiolVenkataraman,thiolatedMolecules}.  However, it is well known that a monolayer of non-functionalized molecules can bind to clean surfaces and be physisorberd by van der Waals forces. These forces may be strong enough to perform electronic transport measurements at ambient conditions, what constitute the starting hypothesis in our study.
In this work, we study the effect of physisorption on gold of different bare molecules often used as organic solvents in molecular electronics: benzene (C$_{6}$H$_{6}$), cyclohexane (C$_{6}$H$_{12}$) and toluene (C$_{6}$H$_{5}$CH$_{3}$). The presence of these molecules on the surfaces is characterized using a Scanning Tunneling Microscope \cite{STMOrigen} (STM), and their typical conductance fingerprint through STM-based Break-Junction (STM-BJ) experiments\cite{Xu2003}, both operated at ambient conditions see  Figure \ref{fig:SchemeTechniques} illustration (a) and (b)).

\section{Methods and Material} 
\begin{figure}
\centering
\includegraphics[width=0.99\linewidth]{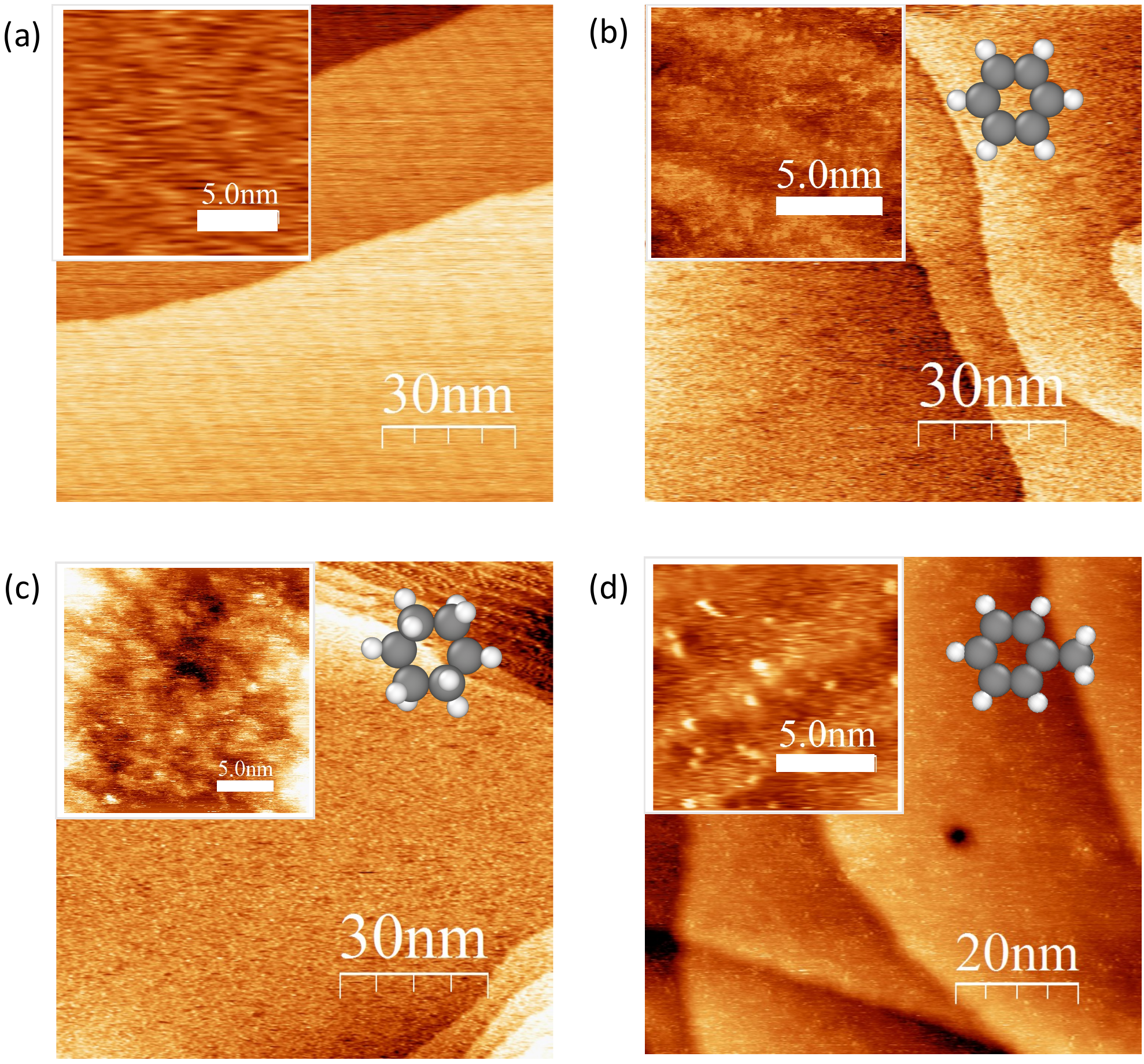}
\caption{Scanning Tunnelling Microscope images of a physisorbed molecular layer on Au(111) (a) clean gold surface (b) benzene (c) cyclohexane (d) toluene. The STM images were performed using tunnelling currents in the range 0.1-1.0 nA,  and a voltage difference in between tip and sample of about 0.3-1.0 V (details in sup. inf. section}
\label{fig:Topography}
\end{figure}

For the experiments a freshly distilled, pre-purified solvent (or compound) was employed for every experiment (details in the purification process at sup. inf. section). These high purity solvents, benzene ($>99.95 \, \%$), cyclohexane ($>99.98 \, \%$) and toluene ($>99.95 \, \%$) were stored in borosilicate glass containers. In the case of STM experiments, gold substrates from Arrandee \cite{Arrandee} were embedded in piranha solution in a ratio 3:1 of sulfuric acid and hydrogen peroxide to clean the surface. Later, for removing the piranha solution, the samples are immersed in miliQ water.  A flame annealing process is performed to acquire flat gold. Imaging of the surface is carried out as to ascertain the cleanness and formation of the (111) terraces of gold with no traces of adsorbed molecules. It should be pointed out that in our images the herringbone reconstruction\cite{Woll1989} is clearly seen indicating the cleanness of our gold surfaces under ambient conditions (see Figure \ref{fig:Topography} (a)). 

Right after the image of clean Au surfaces was obtained, we used two methods for the disposal of the molecules over the surfaces obtaining for both identical results. 
First, we used the drop-casting method in which the solvent was deposited onto the surface and dried under argon gas. Thereafter, the sample was dried-out down to $ 10^{-3}$ mbar of pressure. The second method is based in ambient temperature thermal evaporation,  where the gold substrates were held above the solvent surface (non-immersed) in a closed glass vial. 

\section{Results and discussion}
It should be stressed that prior to the STM imaging of the surfaces, we made our best to evaporate the deposited molecules with the aid of argon gas blowing and vacuum. Our STM images following the procedure above are shown in Figure \ref{fig:Topography}. In our images the Au(111) terraces are still clearly visible indicating that our deposition procedure did not affect the quality of the gold surface. However, even after our efforts to remove the molecules, these decorate the gold surfaces through a reproducible molecular arrangement. 
In Figure \ref{fig:Topography}, STM topographic images show the presence of the three deposited molecules (b) benzene, (c) cyclohexane and (d) toluene, decorating the surfaces. Following the procedure of preparation and evaporation the topographic images show a high reproducibility. The reported results were obtained on the first day of scanning  although similar results could be obtained on the second day. 

Topographic images were acquired using a bias voltage between 0.3 and 0.7 V  applied to the tip in a current constant mode of 0.1-1.0 nA (details in sup. inf. section). 
The inserts in Figure \ref{fig:Topography} show zooms of regions where it is feasible to distinguish single molecules as bright dots. Although atomic resolution of the molecules is not possible due to the molecular thermal diffusion at ambient conditions, the high molecular coverage of the molecules over the surface restricts enough the movement of the molecules allowing us to obtain details of the molecular sizes as shown in Figure \ref{fig:SchemeTechniques}(c). The possible interactions among the molecules and molecules-substrate (mainly van der Waals forces) can affect the geometry binding, albeit when working with planar molecules as we are studying here, the most probable spatial distribution of the molecules would be parallel to the substrate. For the reasons depicted above, the ring structure of the three molecules cannot be differentiated in the images, but the measurement of the molecular diameter is feasible. This procedure is described in the work of Baciu \textit{et al.} \cite{Baciu20} by taking the width at mid-height of the height profile of the molecules (Figure \ref{fig:SchemeTechniques} (c)). The measured sizes are compare with the literature values which are in accordance with them as collected in table \ref{tab:tab1}.

\begin{table}[htbp]
\caption{First column indicates the molecule, second column shows the mean value and standard deviation of our diameter measured experimentally $(\phi_{exp})$ and, third column offers the diameter found in the literature $(\phi_{lit})$.}
    \centering
    \begin{tabular}{ccc}
    \hline
       Molecule & \begin{tabular}[c]{@{}c@{}}  $\phi_{exp}$ [nm]\end{tabular} & \begin{tabular}[c]{@{}c@{}}$\phi_{lit}$ [nm]\end{tabular} \\ \hline
      Benzene     & $0.34\pm 0.01$  & 0.28\cite{IQMOL}, 0.25 \cite{sizeMoleculesB}, 0.38 \cite{sizeMoleculesBC}\\   
      Cyclohexane & $0.40 \pm 0.02$    & 0.31\cite{IQMOL}, 0.49 \cite{sizeMoleculesBC}\\   
      Toluene     & $0.38 \pm 0.01$  & 0.42\cite{IQMOL}  \\  \hline
\end{tabular}
\label{tab:tab1}
\end{table}

Our results clearly confirm the presence of the three organic solvents on the gold surface but only in the thin layer, as the Au(111) terraces are still visible.
This indicates that, as we expected, the very first layers of the three organic solvents under study are firmly physisorbed to the gold surface. 

The existence of strong binding forces in between the gold surface and the studied molecules opens the possibility of forming single metal-molecule-metal junction without the need of the presence of anchoring groups.

For studying the formation of molecular bridges of our three organic solvents we use an improved STM-BJ experimental setup  under room conditions. This technique\cite{Pascual1993,Agrait1993} is, with the  Mechanically Controllable  Break Junction \cite{Krans93} (MCBJ), one of the most common approaches use to study the electronic transport in atomic \cite{Krans96,Agrait2003} or molecular conductors\cite{Venkararaman2013}.
In our STM-BJ  experiments tip and surface electrodes are formed  by two gold wires (0.5 mm, GoodFellow, 99.999\%) facing the curved cylindrical surfaces one against the other in a perpendicular way. In this technique, the two electrodes can move one respect to the other, as depicted in Figure \ref{fig:SchemeTechniques}(b). They are firstly crashed and then, gently retracted allowing for the formation of the molecular bridge (more details about STM-BJ experimental setup  in sup. inf. section). 
During this procedure, the conductance of the molecular bridge is recorded at a constant bias voltage (in our case $\sim$100 mV) and conductance vs. electrode-displacement curves are built, the so-called traces of conductance.  Our results and analysis are based in the traces of conductance during rupture of the bridge (rupture traces) as  shown in Figure \ref{fig:SchemeTechniques} panel (d). 

In the rupture traces, we first observe the breaking of the metallic bridge formed in between the two electrodes in the range above 1 $G_0$. Below this breaking point, we enter into the tunneling regime of conductance which is originated from the presence of tunnel barriers when the electrodes are separated, and can be described by the exponential decay. The presence of a molecular bridge will be shown as a plateau in conductance vs. relative displacement, instead of the exponential decrease. The transmission of the conductance channel through the molecule will depend on many factors including the alignment of the molecular levels to the Fermi energy of the electrodes in such a way that the conductance will get close to $G_0$ for the fully opened channel, otherwise it would decrease exponentially as the energy of the molecular orbitals of the molecular bridge move away from the Fermi energy of the electrodes and the electrodes are being separated.


We have characterized the Au-Molecule-Au bridges for the three molecules under study.
For this experiments we prepared the electrodes to assure the cleanness in break-junctions experiments (see sup. inf section). Once we guaranteed the absence of contaminants, we introduce the target molecules directly on the surface. Thus, when performing the experiment for each molecule, we can observe electrode-molecule junctions plateaus in the rupture traces, as shown in Figure \ref{fig:SchemeTechniques}(d). Our technique allows the recording of these traces at a rate of 100 samples/s for traces containing 1024 points each. From these rupture traces we can construct conductance histograms, which reveals in a peak shape the most frequent values of conductance obtained. These histograms can be expressed in linear scale to obtain information for values bigger than 1 $G_0$ or in semi-logarithmic scale to highlight the values below the quantum of conductance. Every different molecule and their characteristic orientations respect to the electrodes can be distinguishable in the histograms as  visible peaks, converting the histogram in a unique fingerprint of every different single metal-molecule-metal junction \cite{Mishchenko2010}. 

In our study, we can distinguish the traces where a molecule was present from those were only traces of gold single-atom contact, followed by a tunneling current from the electrodes. The main difference in between these two would be the formation of conductance plateaus or not.
In table \ref{tab:percentage} is reported the actual relative traces which contain the electrodes data and the molecular contributions separated by data selection.  For this purpose, we define it as a criterion for considering clean gold when the conductance trace do not have significant counts under 0.8 $G_{0}$. In the literature the characteristic gold peak is in the range between 0.8 and 1.2 $G_{0}$ \cite{goldCriteria,RoleSab18}. 

\begin{table}
\centering
\caption{Percentage of data traces with and without the presence of a molecular bridge.}
\begin{tabular}{cccc}
\hline
             & Benzene & Cyclohexane & Toluene \\ \hline
Gold contact & 50 \% & 20 \% & 30 \% \\ 
Molecular bridge & 50 \% & 80 \% & 70 \% \\  \hline
\end{tabular}
\label{tab:percentage}
\end{table}

The presence of a molecular bridge will depend on different factors such as, the degree of coverage of the different molecules on the surfaces, their diffusion over the surface or the strength of the Au-Molecule binding\cite{Wetterer1998}.
In our results it appears that benzene is not present in 50\% of our junctions while this percentage decreases to 20\% and 30\% in cyclohexane and toluene respectively. 
The low amount of traces of benzene could be in a first thought related for being the one of the three with the lowest boiling point\cite{NIST}, but this dependence does not  hold for the case of the other two molecules. Another possibility is to look at the strength of the activation energy for desorption\cite{Wetterer1998}, but here the lowest value corresponds to the case of cyclohexane. Other possible factor can be attributed to the geometries including the flatness of the molecules. A combination of different factors may explain these differences.

\begin{figure*}
\centering
\includegraphics[width=0.6\linewidth]{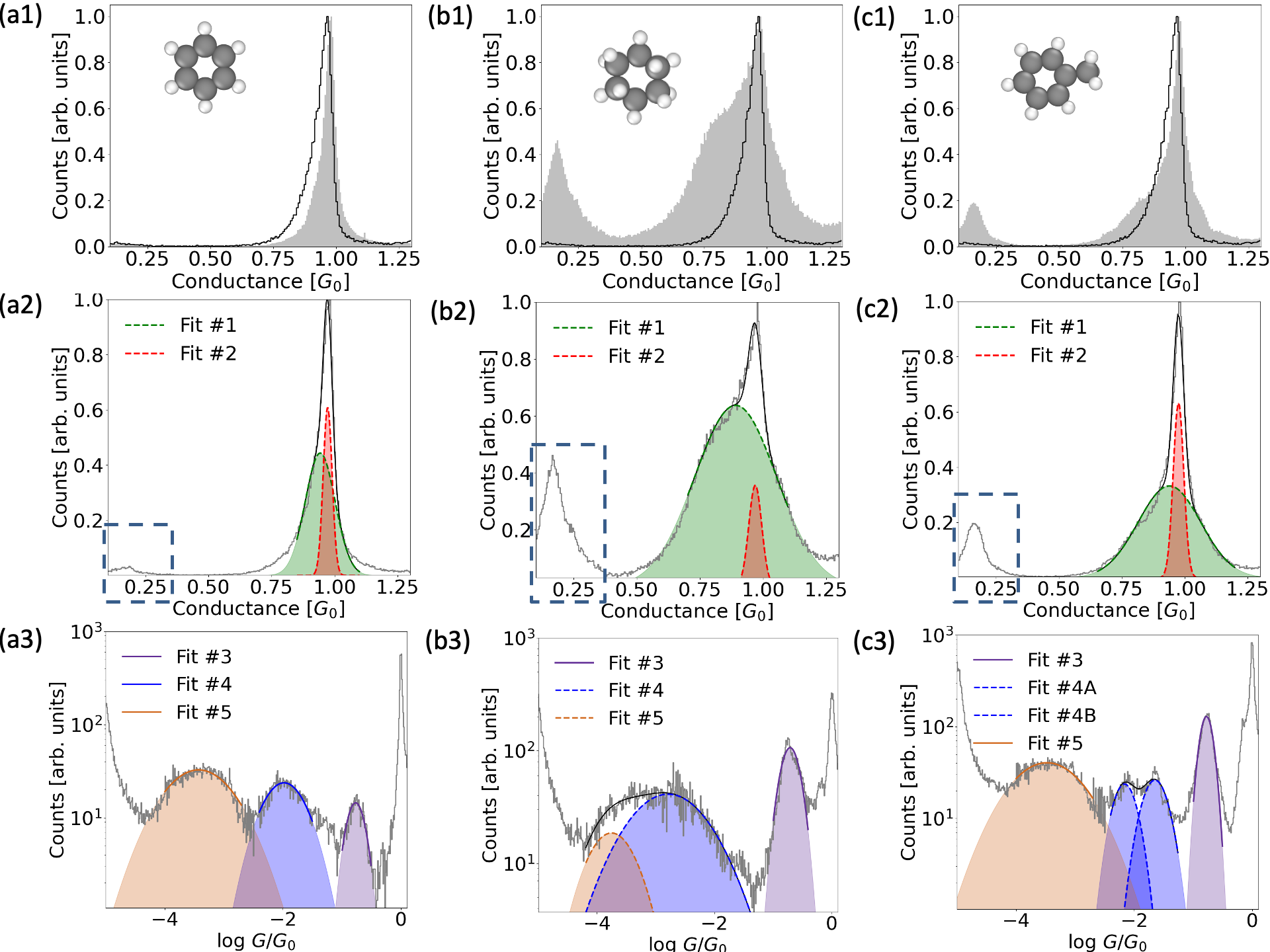}
\caption{Conductance histograms (normalized to the maximum value) of the three solvents deposited on gold. First (a1-a3), second (b1-b3) and third (c1-c3) columns show the results of benzene, cyclohexane and toluene respectively. First row presents a comparison between the clean histogram of gold (black line) with the histogram of conductance obtained when molecules are inserted (shaded gray area). In the second row the conductance of a single-atom gold contact is analyzed by Gaussian fits. Green and red shaded areas may be related to the monomeric and dimeric configurations of gold summed with the molecular contributions (Fit $\#$1 and $\#$2). The molecular contribution is indicated by the rectangle in dash blue. To resolve the molecular peaks at lower conductance value, the histograms are presented in logarithmic scale in the third row. Purple, blue and brown measurements come from the molecules when they adopt different attachments to the electrodes while stretching the junction (Fit $\#$3, $\#$4 and $\#$5). Here the black lines represent the sum of two Gaussian contributions which fits with the data trend. Each shaded colored area is supposed to be related to different binding configurations.}
\label{fig:GoldSolventHist}
\end{figure*}

For the traces in which the molecules were anchored we constructed three different histograms that constitute the fingertip of the molecules, as stated above, and can give us important information on the overlap of the molecular orbitals to electrons at the Fermi energy at the metallic electrodes. 
Figure \ref{fig:GoldSolventHist} displays a collection of histograms built from a set of thousands of rupture traces as previously defined . For the statistical analyses, around 3000 traces of clean gold has been considered for reference. On the other hand, the molecular contribution histograms were constructed from over 10000 traces for the  different molecular junctions.
These histograms are represented in Figure \ref{fig:GoldSolventHist}. 

In order to analyze the statistical distribution, we fit the data to Gaussian functions $\left( f(x) = \sum_{i=1}^{N} A \, e^{{ -({x - \mu })^2 } / {2\sigma ^2}}\right)$, were $\mu$ corresponds to the peak of the most probable conductance. The typical histogram of gold is described in Figure S5 of the supporting information$^\dag$ where the histogram of clean gold is shown.
In Figure \ref{fig:GoldSolventHist} panels (a1),(b1) and (c1) compare the typical histogram of clean gold (black line) with the histogram of the respective molecular contribution (shaded gray area).
The most probable values of conductance are collected in tables \ref{tab:G0conductance} and \ref{tab:lowerconductance} as a function of the regime under study. It is important to note that below the quantum of conductance, we plot the logarithmic data and also, a log scale representation was used to emphasize the peaks.

The molecular contribution is observed in the linear and logarithmic data representation in Figure \ref{fig:GoldSolventHist}. The strength of the molecular peak depends on the amount of bare gold traces and therefore is higher for the case of cyclohexane and toluene.

Linear histograms (Figure \ref{fig:GoldSolventHist} (a1), (b1) and (c1)) show a slight shift of the peak related to the atomic contact of gold at the quantum of conductance (black line). To unveil its origin, we use a two Gaussian fit in this region (Figure \ref{fig:GoldSolventHist} (a2), (b2) and (c2)) following the work of Sabater \textit{et al.} \cite{RoleSab18}. In their manuscript, they characterize the monomeric, dimeric and double contact contributions for gold, silver and cooper histograms of conductance through the fitting of at least three Gaussians. Note that a monomer means the presence of solely one atom attached to the electrodes and a dimer is built by two atoms in line between the electrodes. This procedure allows us to disentangled our peak as the sum of two main contributions related to the formation of monomers (Fit $\#1$) and dimers (Fit $\#2$) with the molecular contribution. The fitting parameters are collected in table \ref{tab:G0conductance}.

\begin{table*}
\centering
\caption{Gaussian fits to the conductance close to 1 $G_0$ for pure gold and the molecular contribution. Color code of the letters matches to the colored areas used in panels (a2), (b2) and (c2) Figure \ref{fig:GoldSolventHist}.}
\begin{tabular}{ccccc}
\hline
\begin{tabular}[c]{@{}c@{}}\end{tabular}&
\begin{tabular}[c]{@{}c@{}}Pure Gold\\ $\mu_{i} \pm \sigma_{i}$\end{tabular} & \begin{tabular}[c]{@{}c@{}}Benzene\\ $\mu_{i} \pm \sigma_{i}$\end{tabular} & \begin{tabular}[c]{@{}c@{}}Cyclohexane\\ $\mu_{i} \pm \sigma_{i}$\end{tabular} & \begin{tabular}[c]{@{}l@{}}Toluene\\ $\mu_{i} \pm \sigma_{i}$\end{tabular} \\ \hline
Monomer & \textcolor{Mygreen}{0.92 $\pm$ 0.06} & \textcolor{Mygreen}{0.94 $\pm$ 0.07} & \textcolor{Mygreen}{0.89 $\pm$ 0.16} & \textcolor{Mygreen}{0.94 $\pm$ 0.12} \\ \hline
Dimer & \textcolor{Myred}{0.96 $\pm$ 0.02} & \textcolor{Myred}{0.97 $\pm$ 0.02} & \textcolor{Myred}{0.97 $\pm$ 0.03} & \textcolor{Myred}{0.98 $\pm$ 0.02} \\ \hline
\end{tabular}
\label{tab:G0conductance}
\end{table*}

One should notice that, although the conductance distribution for the dimeric contacts is unaltered with the presence of the molecules, there is an increase in the width of the distributions for the monomeric contacts when the molecule is present.
This shift may be related to a contribution of the molecules to the conductance. Also, there is a widening in the peak around 0.8 $G_{0}$ in the linear scale, though it is not appreciated for benzene. 

Now, if we focus in the peaks coming from the molecular bridges below the quantum of conductance, we distinguish for each molecule, three main distributions shown  with different colored areas (see Figure \ref{fig:GoldSolventHist} panels (a3), (b3) and (c3)).
It is quite remarkable to notice the strong similarity in between the distributions that may come from the similarity, also, in between the three molecules. Benzene and cyclohexane are prototypes of an aromatic (delocalized $\pi$ electrons in the molecular plane) and aliphatic molecule (localized sigma electrons), while toluene is a mixed aromatic-aliphatic molecule sharing features from both. Therefore one should expect differences in between their conductances to be subtle, similar for toluene and benzene and lower for cyclohexane.
Our data fits only logarithm scale, therefore these fitting values are logarithmic (fitted conductance values ($\mu$) along their deviations ($\sigma$) are collected in Table S5 sup. inf. section). In table \ref{tab:lowerconductance}, we summarize the fitting parameter for the different conductance peaks as a function of $G_0$. 

\begin{table}
\caption{Most probable conductance values (G) obtained from Gaussian fits under the quantum of conductance regime for solvent molecules.  Color code of the letters matches to the colored areas used in Figure \ref{fig:GoldSolventHist} (a3), (b3) and (c3).}
\begin{tabular}{lcccccc}
\hline
{\color[HTML]{FFFFFF} } & $G_{Fit \, \#3}$ {[}$G_{0}${]}                            & $G_{Fit \, \#4}$ {[}$G_{0}${]}                                                & $G_{Fit \, \#5}$ {[}$G_{0}${]}            \\ \hline
Pure Gold                                       & \multicolumn{3}{c}{Featureless}                                                                                                                                                        \\ \hline
Benzene                                         & {\color[HTML]{6200C9} 1.6 $\cdot 10^{-1}$ } & {\color[HTML]{3166FF} 1.1 $\cdot 10^{-2}$}                                                                &  {\color[HTML]{963400} 3.8 $\cdot 10^{-4}$}\\ \hline
Cyclohexane                                     & {\color[HTML]{6200C9} 1.9 $\cdot 10^{-1}$ } & {\color[HTML]{3166FF} 1.5 $\cdot 10^{-3}$}                                                                &  {\color[HTML]{963400} 1.7 $\cdot 10^{-4}$}\\ \hline
Toluene                                         & {\color[HTML]{6200C9} 1.7 $\cdot 10^{-1}$} & {\color[HTML]{3166FF} \begin{tabular}[c]{@{}c@{}}2.3 $\cdot 10^{-2}$\\ 7.1 $\cdot 10^{-3}$\end{tabular}} &  {\color[HTML]{963400} 3.4 $\cdot 10^{-4}$}\\ \hline
\end{tabular}
\label{tab:lowerconductance}
\end{table}

The highest peak in the range $(1.6-1.9) \cdot 10^{-1}$ $G_0$ is almost identical for the three molecules. As these conductances are so similar, the delocalized electrons present in benzene or toluene should not play any role. Therefore, the most probable origin for this peak should be related to the presence of the molecular ring being perpendicular to the direction of the electrodes. We should point out that this result for benzene is in good agreement with the work by Tal's group \cite{Tal2016} at low-temperature using silver electrodes.
The second peak is the one showing the main differences in between the three molecules, differences which follows our initial guess stated above. In this case, the conductance is similar for benzene ($1.1 \cdot 10^{-2}$ $G_0$) and toluene $((0.7-2.3) \cdot 10^{-2}$ $G_0)$ but much lower and difficult to disentangle from the third for the case of cyclohexane ($1.5\cdot10^{-3}$ $G_0$). It is also remarkable to observe a splitting of the second peak for the case of toluene most likely coming from the presence of the methyl group. 
Other peaks at lower values may appear from molecular staking \cite{stackedLatha}.

\begin{figure*}
\centering
\includegraphics[width=0.8\linewidth]{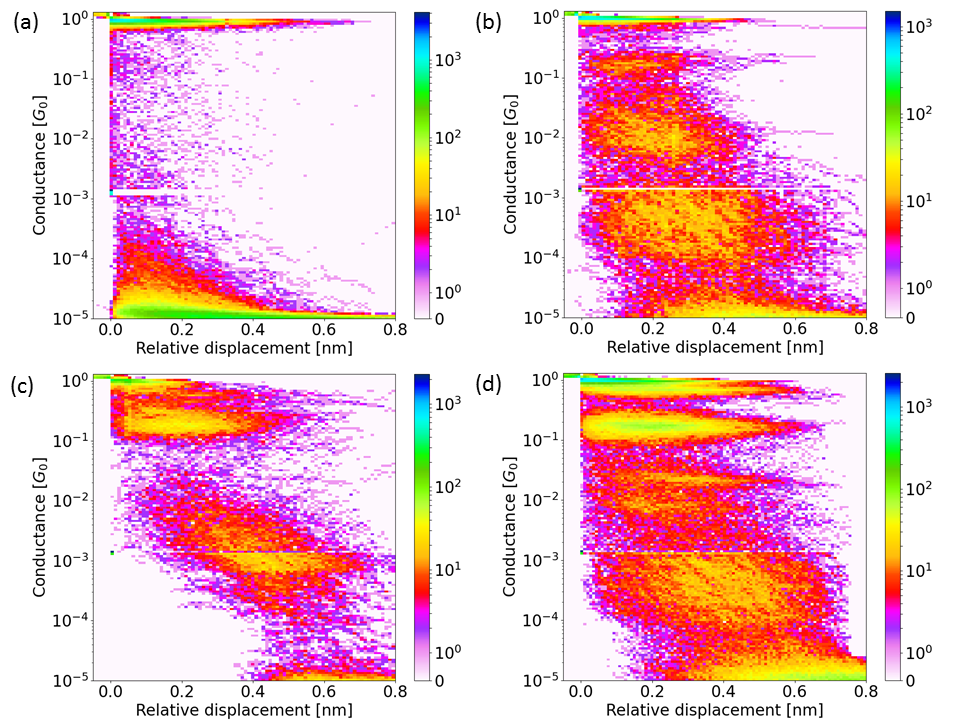}
\caption{Density plots according to the conductance histograms of (a) clean gold, (b) benzene, (c) cyclohexane and (d) toluene.}
\label{fig:2Dplot}
\end{figure*}

To get more insight, we  analyzed the overall shape of the evolution of the conductance traces. For this we study the average trace for each molecule that can be constructed as a 2D density plot as shown in Figure \ref{fig:2Dplot}. For making this plot we took into consideration only the traces with molecular contribution, which are traces with plateaus below the atomic conductance. The point at which conductance takes the value of, at least, the one of the atomic contact of gold (G $<$ 1.1 $G_0$), is taken as the origin for the electrodes relative displacement, and then these are added up to form the 2D density plot. 

The first of the density plots (a) shows that, for the case of clean gold electrodes at ambient conditions, it is feature-less from about 5 $\cdot 10^{-4}$ $G_0$ to the region of the atomic contacts at about $G_0$. Below 10$^{-4}$ $G_0$ we can observe a signal coming from the tunneling current between electrodes that fade to the electrical noise of our amplifier when the electrodes are separated about 1.0 nm. We should notice that the color scale is logarithmic to make noticeable even individual events.   

In these representations the contribution of the molecules is highlighted allowing the different details of the various  geometrical configurations of the molecular bridges to be examined.  The most important information given by this representation is the overall length and shape of the conductance plateaus for the different shapes for the molecular bridge. It is remarkable that the main differences observed are in the second peak contributions ($\sim 10^{-2}$ $G_{0}$), especially in cyclohexane.

\begin{figure}
\centering
\includegraphics[width=0.99\linewidth]{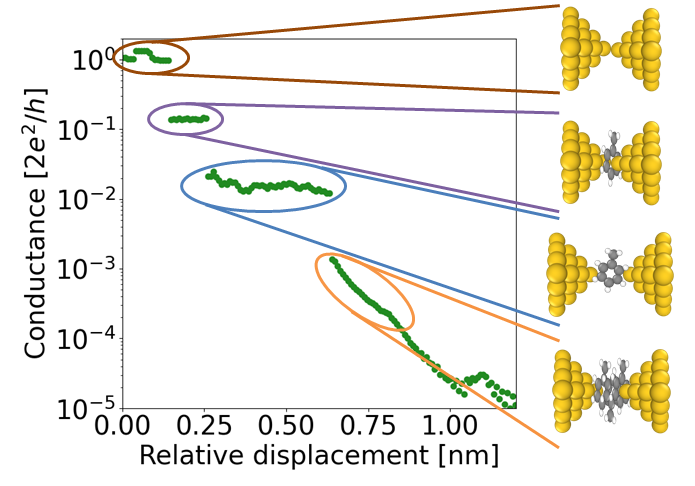}
\caption{A toluene rupture trace with the presence of different plateaus related to different atomic and molecular geometrical configurations. Atomic gold sized-contact, perpendicular, parallel and staked molecular bridges are illustrated from top to the bottom, respectively. Color code of circles and lines matches to the colored areas used in panels (a3), (b3) and (c3) Figure \ref{fig:GoldSolventHist}.}
\label{fig:geometricalConf}
\end{figure}

In Figure \ref{fig:2Dplot} we observe two features that are coincident for the three studied molecules. First, the plateaus at about $10^{-1}$ $G_{0}$ are plateaus remaining quite constant in value while stretched and the length of the plateaus is similar (slightly shorter) to the one related to the one-atom bridge between electrodes (traces at $\sim$ 1 $G_{0}$). This indicates that in this configuration  the bind of the molecules to the electrodes is similar (slightly lower) to the one of the one atom. As we concluded from the histograms before, all the facts imply that these traces are most likely related to the plane of molecules being perpendicular to the direction of the contact (see Figure \ref{fig:geometricalConf}). 

The second set of traces shared by the three molecules is the one with a mean conductance of about $10^{-3}$ $G_{0}$. In this occasion the plateaus are longer and decrease over 1-2 orders of magnitude in conductance. This fact implies a weak bonding of the molecules to the electrodes. This configuration is also shown to happen, on average, $\sim$0.2 nm after the gold contact was formed, normally even after the presence of  a previous molecular junction. The conductance plateaus in this region are also much longer than the ones related to the atomic contact. The three features above, long plateaus after the atomic and molecular contacts and strong decrease of conductance while stretching, indicate a weak junction most likely related to the conductance over a number of staked molecules\cite{stackedLatha} (see Figure \ref{fig:geometricalConf}). The fact of a thin molecular layer over the metallic surface may limit the number of staked molecules and may be the reason why there is a lower value for that peak at about $3 \cdot 10^{-5}$ $G_{0}$.

Where the differences in between the three molecules unveil is on the set of traces that for the case of benzene or toluene can be found at about  $10^{-2}$ $G_{0}$. In this case, we find again a set of traces with plateaus of constant conductance and with plateau lengths similar to the ones related to the one-atom bridge between electrodes. As we previously discussed these facts indicate a well bonded molecular bridge. The variations observed in this peak for the different molecules is an indication of the role of the extended molecular $\pi$ orbital of the molecules in the case of benzene and toluene. Therefore this configuration should correspond to the molecular ring being parallel to the direction of the electrodes (see Figure \ref{fig:geometricalConf}). In fact, for the  case of cyclohexane, although we could fit the histogram in the area of $10^{-3}$ $G_{0}$ to two Gaussian curves, we find difficult to find signatures of this configuration in the density plot. It is possible that we are not detecting the signatures of this geometrical configuration for cyclohexane due to the lack of an extended molecular orbital in this molecule.  Finally, we should notice that in the case of toluene this peak is split, most likely, due to the presence of the methyl group which breaks its symmetry. This can be taken as a fingerprint to distinguish in between aliphatic and aromatic series of molecules.
\section{Conclusions}
In summary, we have proven the presence of an adsorbed molecular layer over a gold surface, of the three organic solvents, benzene, cyclohexane and toluene, at ambient conditions. The strength of the adsorbed molecular layer is tight enough to the electrodes to permit the characterization of their role in molecular electronics. 

The presence of the molecules on Au(111) surfaces has been clearly identified through STM topographic images. We have been able to distinguish individual molecules of sizes between 3 and 4 \AA. The observation of thin molecular films decorating the gold substrates, supports our conclusion that our measurements are related to stable single-molecules junctions. The study of the conductance characteristics of the molecular junctions has allowed us to identify the different conductive states of single molecules attached to gold electrodes and to establish their characteristic conductance signature.

Our results imply the existence of an important contribution of the studied molecules on surfaces at ambient conditions. Their influence should be not neglected when these are being used for the preparation of experiments at ambient conditions as we have proven the difficulty of their removal through common preparation procedures and their role in different experiments in molecular electronics.

\section*{Conflicts of interest}
There are no conflicts to declare.

\section*{Acknowledgements}
This work was supported by the Spanish Goverment (MAT2016-78625-C2 and PID2019-109539GB-C41) and the Generalitat Valenciana through PROMETEO/2017/139 and program CDEIGENT/2018/028.
We want to thank to Prof. N. Agra\"it his fruitfully tips and advises in the use of combined I-V converters.

\section{Supporting Information}
\subsection{Experimental details  STM Topography}

(a) The topographic image of a clean and flat gold surface were taken at $I_{t}$ = 0.91 nA and $V_{bias}$ = -0.3 V. (b) Main topographic image of benzene with parameters $I_{t}$ = 0.83 nA and $V_{bias}$ = -0.3 V. Inset image bias parameters are $I_{t}$ = 0.83 nA and $V_{bias}$ = -0.5 V. (c) Topographic image of cyclohexane with parameters $I_{t}$ = 0.11 nA and $V_{bias}$ = 0.9 V. Inset image bias parameters are $I_{t}$ = 0.39 nA and $V_{bias}$ = 0.4 V. (d) Topographic image of toluene with parameters $I_{t}$ = 0.32 nA and $V_{bias}$ = -0.7 V. Inset image bias parameters are $I_{t}$ = 0.30 nA and $V_{bias}$ = -0.6 V.

\subsection{Experimental details STMB-BJ}
\subsubsection{Preparing the electrodes for STM-BJ}

For STM-BJ experiments, fresh gold wires were scraped in order to verify their cleanness and discard the presence of contaminants. When the cleanness process was ultimately necessary, samples were embedded in hot acetone, ethanol and isopropanol (IPA) while ultrasonicated, followed by a dried step, using argon gas. After, the substrates were treated under argon and oxygen plasma. These processes allow us to discard the possible contaminants deposited on the surface. At last, the surfaces were scraped.
  
\subsubsection{Description of the  STM-BJ experimental setup}
One electrode of our setup is connected to a constant bias voltage $(V_{bias})$ source, whereas the other electrode is connected in series to an IV converter which amplifies the current ($10^6$ V/A). The output of this amplifier is split into two channels; first channel is read in an input of an Analogical Digital Converter(ADC) system and the second channel is connected in a series with a 100 $\Omega$ resistance which is also connected in series with a second IV converter that amplifies 100 times the first signal. The output of the second IV is connected to another input channel of the ADC system. A homemade software based in LabView code merge both current converted signals ($I_{merged}$) that cross the atomic/molecular conductor. Alongside, the relative displacement of the tip and the current ($I_{merged}$) is recorded during the indentation and retraction process. Knowing that in our case $V_{bias} = 100 mV$, the conductance can be expressed  $(G=V_{bias}/I_{merged})$ and converted to units of quantum of conductance \cite{VanWees1988} $G_0 = 2e^2/h$, where $e$ and $h$ are the charge of the electron and the Planck constant respectively, and the factor two comes from spin degeneracy. The conductance $G$ can be easily converted in this units  knowing that $G_{0}=\frac{1}{12906} \Omega^{-1}$. Historically the curves conductance vs the relative displacement are named traces of conductance and they can be classified into traces of rupture or formation, depending respectively if the molecular bridge is stretched or compressed.

\subsection{Gaussian fitted parameters under 1 $G_0$}
\begin{figure}
\centering
\includegraphics[width=0.99\linewidth]{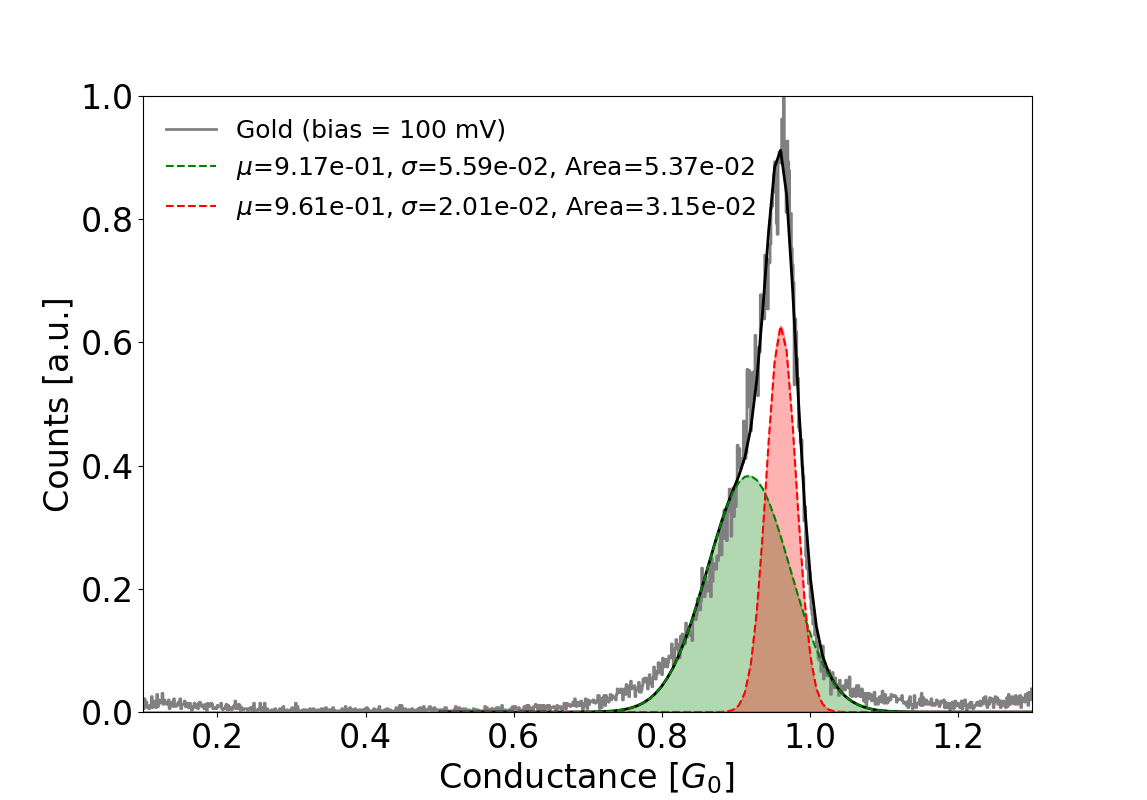}
\renewcommand\thefigure{S\arabic{figure}}  
\caption{Gaussian fit to a set of traces of clean gold.}
\label{fig:GausianFitAU100mv}
\end{figure}

In figure \ref{fig:GausianFitAU100mv} is show the histogram of conductance of clean gold. For the peak around of $1 G_0$ we have fitted two Gaussians. The  mean values ($\mu$),the standard deviations ($\sigma$) and the area under the curve are detailed in the legend of the figure.

\begin{table*}
\centering
\renewcommand\thetable{S\arabic{table}}  
\caption{Logarithmic fitted conductance values ($\mu$) and their logarithmic deviations ($\sigma$) for the solvent molecules.  Color code of the letters matches to the colored areas used in Figure 3 panels (a3), (b3) and (c3).}

\begin{tabular}{c|c|c|c}

\hline

Pure Gold & Benzene & Cyclohexane & Toluene \\ \hline

            & \color[HTML]{6200C9}{-0.8 $\pm$ 0.2} & \color[HTML]{6200C9}{-0.7 $\pm$ 0.2} & \color[HTML]{6200C9}{-0.8 $\pm$ 0.1} \\
Featureless & \color[HTML]{3166FF}{-2.0 $\pm$ 0.4} & \color[HTML]{3166FF} {-2.8 $\pm$ 0.7} & \color[HTML]{3166FF}{-1.6 $\pm$ 0.2, -2.2 $\pm$ 0.2} \\
            & \color[HTML]{963400}{-3.4 $\pm$ 0.6} & \color[HTML]{963400}{-3.8 $\pm$ 0.4} & \color[HTML]{963400}{-3.5 $\pm$ 0.7} \\ \hline
\end{tabular}
\label{tab:lowerconductance}
\end{table*}

\subsection{Purification Solvents} \label{Purification}
\begin{figure}
\centering
\includegraphics[width=0.99\linewidth]{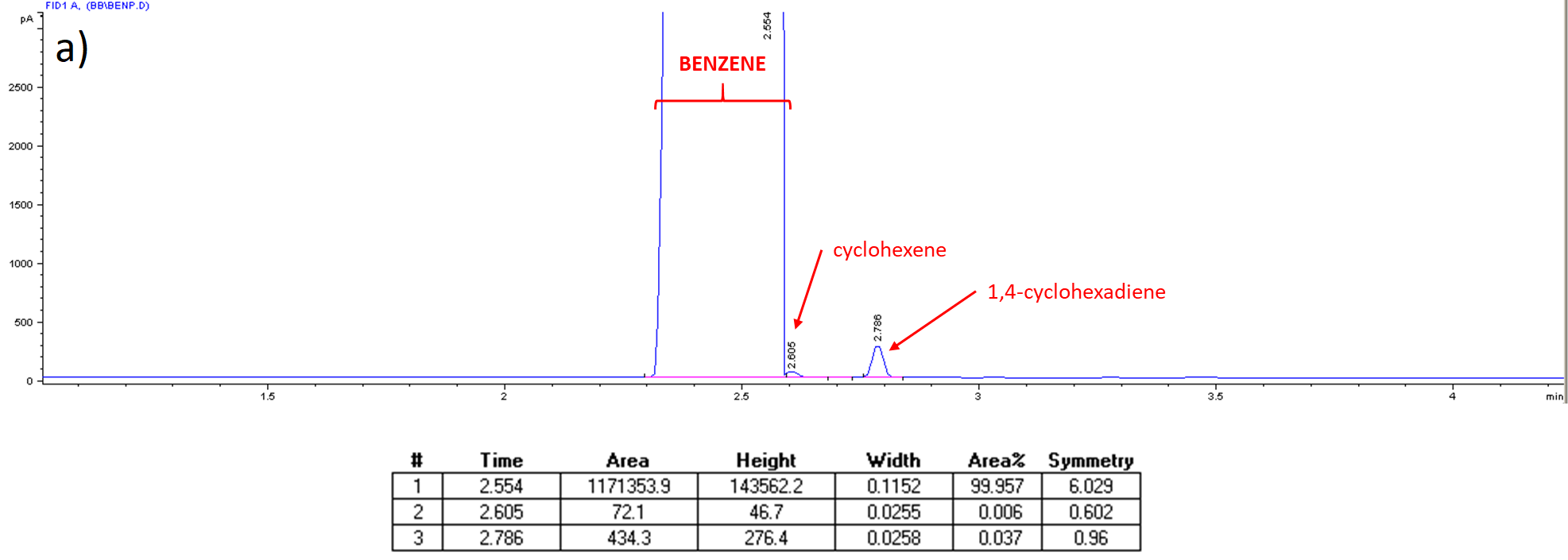}
\includegraphics[width=0.99\linewidth]{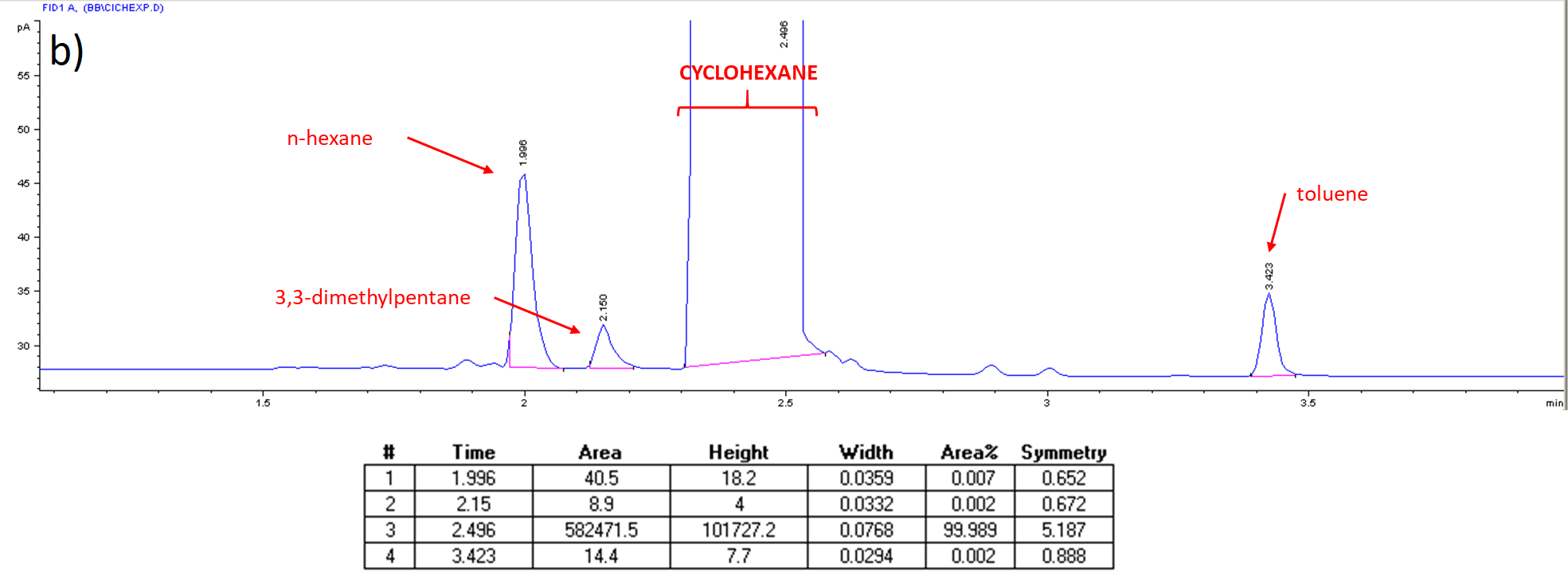}
\includegraphics[width=0.99\linewidth]{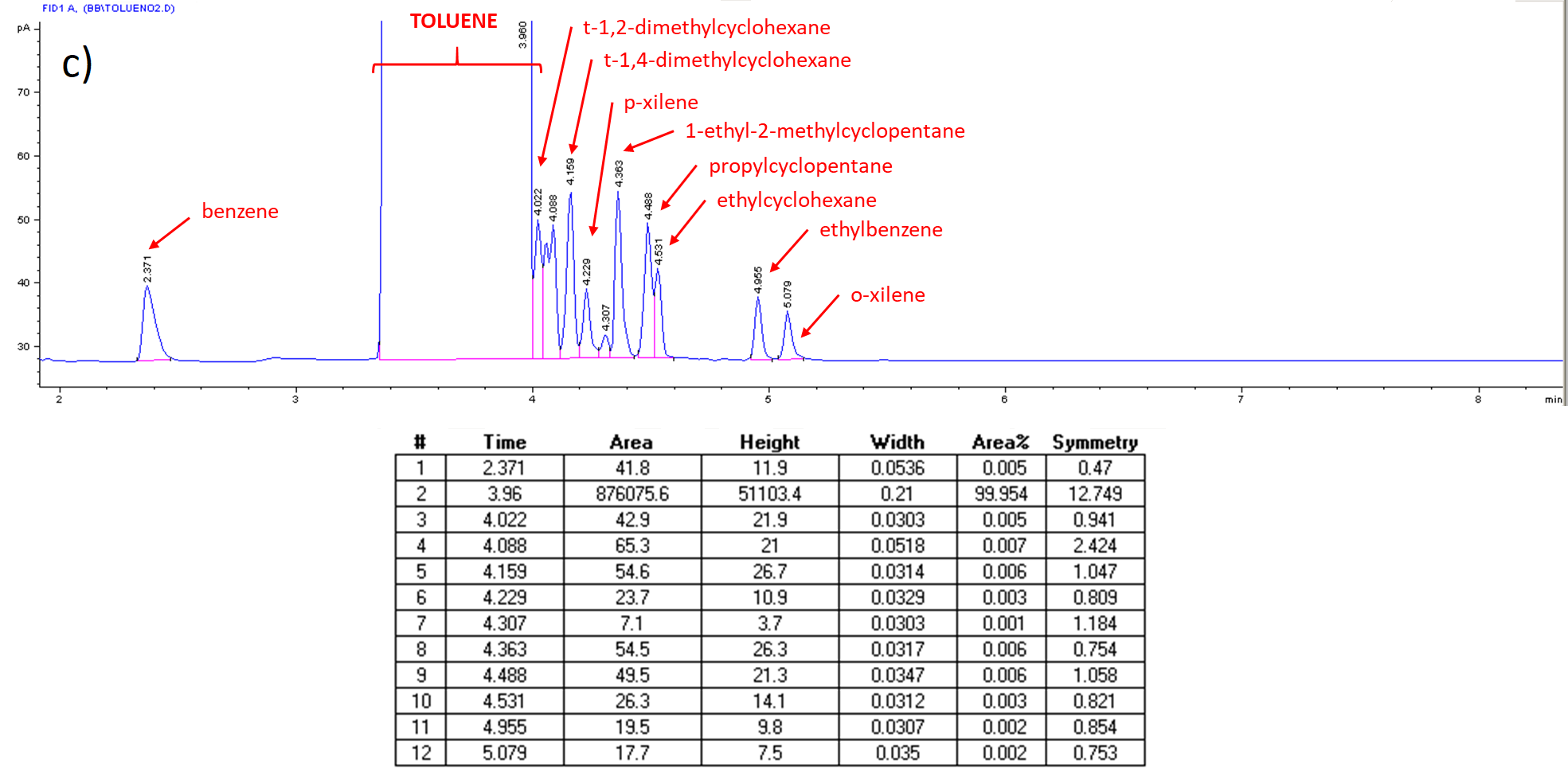}
\renewcommand\thefigure{S\arabic{figure}}  
\caption{From top to bottom, GLC analysis of purified  a) benzene ($>$99.95\%), b) cyclohexane ($>$99.98\%) and  c) toluene ($>$99.95\%) displaying their main constituent trace impurities}
\label{fig:Purity}
\end{figure}

High grade commercially available solvents (benzene\footnote{Caution: Benzene is a well-established cause of cancer to humans. The International Agency for Research on Cancer (IARC) has classified benzene as carcinogenic to humans (Group 1). IARC Monographs on the Carcinogenicity of Chemicals to Humans, Supplement 7; http://www.inchem.org/documents/iarc/suppl7/benzene.html}  for analysis ACS $>$ 99.5\% ITW Reagents; cyclohexane for HPLC $>$ 99.5\% VWR; toluene for analysis ACS $>$ 99.8\% ITW Reagents) were further purified and dried right before use as follows. Benzene, cyclohexane and toluene were passed through a 44\% sulfuric acid-impregnated silica gel chromatographic column at (or below) room temperature. Benzene (mp: 5$^{\circ}$C) and cyclohexane (mp: 6$^{\circ}$C ) were further purified by crystallization under partial freezing conditions. Strictly dry solvents were then obtained by refluxing (ca. 3h) and distilling from sodium-potassium alloy ($K_{2}Na$) under Ar athmosphere. $>$99.95\% Benzene, containing cyclohexane ($<$10 ppm) and 1,4-cyclohexadiene ($<$ 40 ppm), and $>$99.98\% cyclohexane, containing n-hexane ($<$70 ppm),  3,3-dimethylpentane ($<$20 ppm) and toluene ($<$ 30 ppm) as the main detected impurities were obtained in this way. For toluene, a more complex pattern of hydrocarbon impurities was found, consisting of benzene ($<50$ ppm), t-1,2-dimethylcyclohexane ($<50$ ppm), t-1,4-dimethylcyclohexane ($<$60 ppm), p-xylene ($<30$ ppm), 1-ethyl-2-methylcyclopentane ($<$60 ppm), propylcyclopentane ($<$60 ppm), ethylcyclohexane ($<$30 ppm), ethylbenzene ($<$20 ppm) and o-xylene ($<$20 ppm), among others, which did not prevent an overall estimated purity of $>$99.95\%  (Figure \ref{fig:Purity}). Identification of the contaminants was done by GC-MS by matching the MS spectra against the database Wiley275 spectral library. Quantification of impurities was done using GLC provided with a FID detector by spiking the purified solvent with a careful weighted amount of the pure contaminant. Purified solvents were manipulated using borosilicate glass vials and syringes provided with ground glass stoppers and joins, and stainless steel needles under Ar atmosphere. Contact with plastic materials should be avoided at all times.

GC-MS analyses were performed using the electron impact (EI) mode at 70 eV in an AGILENT 5973N mass spectrometer coupled with an AGILENT 6890N gas chromatographer. Gas-liquid chromatography (GLC) quantifications were carried out with a Hewlett Packard HP-5890 instrument equipped with a flame ionization detector (FID) and a 30 m HP-5 capillary column (0.32 mm diam, 0.25 $\mu$m film thickness), using nitrogen as carrier gas (12 psi). Column chromatography was performed with Merck silica gel 60 (0.040-0.063 $\mu$m, 240-400 mesh) impregnated with sulfuric acid (98\%) in a 56:44 (SiO2:H2SO4) ratio. $K_{2}Na$ was prepared by melting potassium (98\% chunks in mineral oil) and sodium (99\% cubes in mineral oil) both from Sigma-Aldrich under Ar atmosphere. 



\bibliography{references} 
\bibliographystyle{rsc} 

\end{document}